\documentstyle[aps,prl,epsf,twocolumn]{revtex}
\begin{document}
\def\bea{\begin{eqnarray}}
\def\eea{\end{eqnarray}}
\def\a{\alpha}
\def\d{\delta}
\def\p{\partial} 
\def\nn{\nonumber}
\def\r{\rho}
\def\rv{\bar{r}}
\def\la{\langle}
\def\ra{\rangle}
\def\e{\epsilon}
\def\o{\omega}
\def\n{\eta}
\def\g{\gamma}
\def\break#1{\pagebreak \vspace*{#1}}
\def\f{\frac}
\def\dg{\dagger}
\def\zh{\hat{Z}}
\twocolumn[\hsize\textwidth\columnwidth\hsize\csname
@twocolumnfalse\endcsname 
\draft
\title{Heat transport in disordered quantum harmonic oscillator chains}
\author{Abhishek Dhar}
\address{ Department of Physics, University of California, Santa Cruz, 
  CA 95064 and \\
Theoretical Physics Group, Raman Research Institute,
Bangalore 560080, India.}
\date{\today}
\maketitle
\widetext
\begin{abstract}
We study heat conduction in quantum disordered harmonic chains connected to
general heat reservoirs which are modeled as infinite collection of
oscillators. Formal exact expressions for the thermal current are
obtained and it is shown that, in some special cases, they reduce to
Landauer-like forms. 
The asymptotic system size dependence of the current is
analysed and is found to be similar to the classical case. It is a
power law dependence and the power depends on the spectral properties
of the reservoirs. 
\end{abstract}

]
\narrowtext

In an earlier paper \cite{dhar}we studied the problem of heat conduction in a
classical disordered harmonic chain that is connected to heat baths specified
by general spectral functions. In this paper we use the
formalism, developed by Ford, Kac, and Mazur \cite{ford} (FKM) to model quantum
mechanical heat reservoirs, 
and extend our earlier calculations to the quantum mechanical regime. Our
results should be useful in  interpreting recent experiments
\cite{schwab} on heat
transport in insulating nanowires and nanotubes. They are also of
interest in the context of the question of validity of Fourier's law in
one-dimensional systems, a problem that has received much attention
recently \cite{four}.   

As pointed out in an earlier paper \cite{adbss}, on transport in
Fermionic wires, the FKM method seems
to be a powerful and relatively straightforward method for studying
nonequilirium transport. Here we use this formalism  to obtain exact
results for the heat current in a disordered harmonic chain. Landauer's
result, suitably generalized to the phononic case, is obtained as a
special case. For general reservoirs the conductance of a wire depends on
details of the reservoirs and contacts.     
Our analysis here closely follows the one followed in
\cite{dhar,conn,rubin} for the case of classical oscillator
chains. The
results obtained in the classical oscillator case and the present
problem have a lot of formal similarity which of course
is because both the problems are linear.    
There has been some earlier work on quantum wires
\cite{zurcher,saito} which follow a
similar approach but we give a more clear and complete picture and
make some interesting predictions for experiments.     

We consider a mass disordered harmonic chain containing $N$ particles
with the following Hamiltonian:
\bea
H=\sum_{l=1}^N \frac{p_l^2}{2 m_l} +\sum_{l=1}^{N-1}
\frac{(x_l-x_{l+1})^2}{2}  +\f{( x_1^2+ x_N^2)}{2} \label{sysH}
\eea
where $\{x_l\}$ and $\{p_l\}$ are the displacement and momentum
operators of the particles  and  $\{m_l\}$ are the random masses.
Sites $1$ and $N$ are connected to two heat reservoirs ($L$ and $R$) which we now
specify. We model each reservoir by a collection of $M$
oscillators. Thus the left reservoir has the following Hamiltonian:
\bea
H_{L} &=& \sum_{l=1}^M \frac{P_l^2}{2 } + \sum_{l , m} \f{1}{2}K_{lm} X_l 
X_m  \label{resH} \\
&=& \sum_{s=1}^M \frac{\tilde{P}_s^2}{2}+ \f{\o_s^2}{2} \tilde{X}_s^2
\nn = \sum_{s=1}^M (n_s+1/2) \o_s a_s^\dagger a_s \nn,
\eea
where $K_{lm}$ is a general symmetric matrix for the spring couplings,
$\{X_l,P_l\}$ are the bath operators  and 
$\{\tilde{X}_l,\tilde{P}_l\}$  are the corresponding normal mode
operators. 
They are related by the transformation $X_l= \sum_s U_{ls}
\tilde{X}_s$ where  $U_{ls}$, chosen to be real, satisfies the
eigenvalue equation $\sum_l 
K_{nl} U_{ls} =\o_s^2 U_{ns}$ for $s=1,2...M$. The annihilation and
creation operators $a_s$, $a^{\dagger}_s$ are given by
$a_s=(\tilde{P}_s-i \o_s \tilde{X}_s)/(2 \o_s )^{1/2}$, etc. and
$n_s=a^\dagger_s a_s$ is the number operator.

The two reservoirs are initially in thermal equilibrium at
temperatures $T_L$ and $T_R$. At time $t=\tau$ the system, which is
in an arbitrary initial state is connected to the reservoirs. We
consider the case where site $1$ on the system is connected to $X_p$
on the left  reservoir while $N$ is connected to $X_{p'}$ on the right
reservoir. Thereafter the whole system evolves through the combined
Hamiltonian: 
\bea
H_T=H+H_L+H_R-k x_1 X_p - k' x_N X_{p'}. \nn
\eea

The Heisenberg equations of motion of the system variables 
are the following (for $t > \tau$):
\bea
m_1 \ddot{x}_1 &=& -[2 x_1-x_2] + k X_p \nn \\
m_l \ddot{x}_l &=& -(-x_{l-1}+2 x_l-x_{l+1}) ~~~1<l<N \nn \\
m_N \ddot{x}_N &=& -[-x_{N-1}+2 x_N]+k' X_{p'}.  
\label{seqm}
\eea
We note that they involve the bath variables $X_{p,p'}$. However
these can be eliminated and replaced by effective noise and
dissipative terms,  by using the equations of motion of the bath
variables. 
Consider  the equation of motion of the left bath variables. They
have the form:
\bea
\ddot{X}_n=-K_{nl} X_l ~~n \neq p \nn \\
\ddot{X}_p=-K_{pl} X_l +k x_1 \label{reseqm}
\eea
This is a linear inhomogeneous set of equations with the solution
\bea
X_n &=& \sum_l [F_{nl}(t-\tau) X_l(\tau)+ G_{nl}(t-\tau)
\dot{X}_l(\tau) \nn \\
&& +\int_\tau^{\infty} dt' G_{np}(t-t') k x_1(t') ~~{\rm where} \label{bsol} \\
F_{nl}(t)&=&\theta(t) \sum_s U_{ns} U_{ls} \cos(\o_s t);\nn \\
~G_{nl}(t) &=& \theta(t) \sum_s U_{ns}
U_{ls} \f{\sin(\o_s t)}{\o_s}. \nn 
\eea
Thus we find that $X_p$ (say) appearing in Eq.~(\ref{seqm})
has the form $X_p(t)= h(t) + k \int_{\tau}^{\infty} dt'
G_{pp}(t-t') x_1(t') $. The first part, given by  $h(t)=\sum_l [F_{pl}(t-\tau)
X_l(\tau)+ G_{pl}(t-\tau)\dot{X}_l(\tau) $, is like a noise term while
the second part  is like
dissipation. The noise statistics is easily obtained using the fact
that at time $t=\tau$ the bath is in thermal equilibrium and
the normal modes satisfy $\la a^\dagger _s(\tau) a_{s'}(\tau) \ra =f(\o_s,\beta_L) \d_{ss'}$.
 Here $f=1/(e^{\beta \o}-1)$ is the equilibrium phonon
distribution and $\la \hat{O}\ra=Tr[\hat{\r} \hat{O}] $ where
$\hat{\r}$ is the reservoir density matrix and $Tr$ is over the
reservoir degrees of freedom. We define the Fourier transforms:
$x_l(\o)=\f{1}{2\pi} \int_{-\infty}^{\infty} dt x_l(t) e^{i \o t}$,
$G^{+}_{pp}(\o)= \int_{-\infty}^{\infty} dt G_{pp}(t) e^{i \o t}$ and $h(\o)=
\f{1}{2 \pi} \int_{-\infty}^{\infty} dt h(t) e^{i \o t}$. Taking limits $M \to \infty$ and $
\tau \to -\infty$ we get:
\bea
&& X_p(\o)=h(\o)+k G^+_{pp}(\o) x_1(\o) \label{noise} \\
&&\la h(\o) h(\o') \ra = I(\o) \d (\o+\o') \nn ~~~{\rm where} \nn \\
&&I(\o)=\f{f(\o) b(\o)}{\pi} \nn \\
&&G^+_{pp}(\o)=\sum_s \f{U^2_{ps}}{\o^2_s-\o^2}-i b(\o)~~~ {\rm and}\nn \\
&&b(\o)=\sum_s \f{\pi U^2_{ps}}{2 \o_s} [\d (\o-\o_s) -\d (\o+\o_s)]. \nn
\eea 
Similarly for the right reservoir we get
$X_{p'}=h'(\o)+k' G^+_{p'p'}(\o) x_N(\o)$, the
noise statistics of $h'(\o)$ being now determined by $\beta'$. The
left and right reservoirs are independent so that $\la h(\o)
h'(\o') \ra = 0$.  
We can now obtain the particular solution of Eq.~(\ref{seqm}) by
taking Fourier transforms and plugging in the forms of $h(\o)$ and
$h'(\o)$. We then get:
\bea
x_l(t) &=& \int_{-\infty}^{\infty} \zh^{-1}_{lm}(\o) h_m (\o) e^{i \o t}
\label{sol} \\
\zh &=& \hat{\phi}_{lm}-\hat{A}_{lm} ~~~{\rm with} \nn \\
\hat{\phi}_{lm} &=& -(\d_{l,m+1}+\d_{l,m-1})+(2-m_l \o^2) \d_{l,m} \nn \\
\hat{A}_{lm} &=& \d_{l,m} [k^2 G^+_{pp}(\o) \d_{l,1}+k'^2
G^+_{p'p'}(\o) \d_{l,N} ] \nn \\
h_l(\o) &=& k h(\o) \d_{l,1}+ k' h' (\o) \d_{l,N}.  \nn
\eea
We can now proceed to calculate steady state values of observables of
interest such as the heat current and the temperature profile. 
We first need to find the appropriate operators corresponding to
these. To find
the current operator $\hat{j}$ we first define the local energy
density $u_l =\f{p_l^2}{4 m_l}+ \f{p_{l+1}^2}{4 m_{l+1}}
+\f{1}{2}(x_l-x_{l+1})^2$. Using the current conservation
equation $\p \hat{u}/\p t+ \p \hat{j}/\p x =0$ and the equations of
motion we then find that $\hat{j}_l=
( \dot{x}_l x_{l-1} +x_{l-1} \dot{x}_l )/2$.  
The  steady state current can now be
computed by using the explicit solution in Eq.~(\ref{sol}). We get
\bea
\la \hat{j}_l \ra =\int_{-\infty}^{\infty} d \o (i \o) [ k^2 \zh^{-1}_{l,1}(\o)
\zh^{-1}_{l-1,1}(-\o) I(\o)  \nn \\+ k'^2 \zh^{-1}_{l,N}(\o)
\zh^{-1}_{l-1,N}(-\o) I'(\o) ] 
\eea
The matrix $Z$ is tridiagonal and using some of its special properties
\cite{adbss} we can reduce the current expression to the following
simple form:
\bea
\la \hat{j}_l \ra &=& \f{ k^2 k'^2}{\pi}  \int_{-\infty}^{\infty} d \o
\f{\o b (\o) b' (\o)}{\mid Y_{1,N} \mid ^2 } ( f-f') \nn \\ &=&
\int_{-\infty}^{\infty} 
d \o J(\o) ( f-f')
\label{jsim}
\eea
where $J(\o)= k^2 k'^2 \o b (\o) b' (\o)/{ \pi \mid Y_{1,N}
  \mid ^2 } $ has the physical interpretation as the total heat 
current in the wire due to all right-moving (or left-moving) scattering
states of the full Hamiltonian (system $+$ reservoirs). Such
scattering states can 
be obtained by evolving initial unperturbed states of the reservoirs
with the full Hamiltonian \cite{adbss,todo}. We
have denoted by $Y_{l,m}$ the determinant of the 
submatrix of $\zh$ beginning with the $l$th row and column and ending
with the $m$th row and column. Similarly let $D_{l,m}$ denote the determinant
of the submatrix formed from $\hat{\Phi}$.

For the special case when the reservoirs are also one dimensional
chains with nearest neighbor spring constants $K_{lm}=1$ and the
coupling constants $k,k'$ are set to unity, we have:
$G^+_{pp}=G^+_{p'p'}=e^{-ik}$ where $\o=2 \sin(k/2)$, $I(\o)=f(\o)
\sin(k)/\pi$ for $ \mid \o \mid  <2$ and $I(\o)=0$ for $\mid \o \mid >
2$. In this case  
Eq.~(\ref{jsim})  
simplifies further and has an interpretation in terms of transmission
coefficients of plane waves across the disordered system. We get:
\bea
J &=& \frac{1}{4 \pi} \int_{-2}^{2} d \o  \o |t_N(\o)|^2
(f-f') ~~~~{\rm{where}} \label{curr} \\
|t_N(\o)|^2 &=& \f{4 \sin^2(k)}{|D_{1,N_s}-e^{ik}(D_{2,N_s}+D_{1,N_s-1})+e^{i 2
k}D_{2,N_s-1}|^2}   \nn 
\eea
is the transmission coefficient at frequency $\o$. We have thus
obtained the Landauer formula \cite{land} for phononic transport. It is only in
this special case of a one-dimensional reservoir and perfect contacts
that we get the 
Landauer formula. The reason is that, only in this case is the  
transmission through the contacts perfect, and this requirement is one of
the crucial assumptions in the Landauer derivation. 
Note that in Eq.~(\ref{curr}) (i)the transmission coefficient {\it does
  not} depend on bath properties and (ii) transmission is only through
propagating modes. For general reservoirs where we need to use
Eq.~(\ref{jsim}) the factor $J(\o)$ involves not just the properties
of the wire but also the details of the spectral functions of the
reservoirs. Thus the conductivity of a sample can show rather
remarkable dependence on reservoir properties as we shall see below.  
The above Landauer-like-formula has earlier been stated in \cite{rego}
and derived 
more systematically in \cite{blen}. 
We note that in the high temperature limit $T~,T' \to \infty$
Eq.~(\ref{curr}) reduces to the classical limit obtained exactly in
\cite{dhar,conn,rubin}.  

{\it Asymptotic system-size dependences :} In the case of electrical
conduction the conductance of a long disordered chain decays
exponentially with system size as a result of localization of
states. In the case of 
phonons the long wavelength modes are not localized and can
carry current. This leads to power-law dependences of the current on
system size as has been found earlier in the context of heat
conduction in classical oscillator chains. A surprising result is that
the conductivity of such disordered chains depend not just on the
properties of the chain itself but also on those of the reservoirs to
which it is connected. It can be shown \cite{dhar} that the asymptotic
properties 
of the integral in Eq.~(\ref{jsim}) depend on the low frequency ($\o
\lesssim 1/N^{1/2}$) properties of the integrand. This means that we
will get the same behaviour as in the classical case. The details can
be found in \cite{dhar,conn,rubin}. Here we summarize
some of the main results:

(i) For the one-dimensional reservoir, which corresponds to the
    classical Rubin-Greer model \cite{rubin}, we get $J \sim \f{1}{N^{1/2}}$.

(ii) The case of reservoirs which give delta-correlated Langevin noise
correspond to the classical Lebowitz model \cite{cash,conn} where one
gets $J \sim \f{1}{N^{3/2}}$.

(iii) In general one gets $J \sim \f{1}{N^{\alpha}}$ where $\alpha$
depends on the low frequency behaviour of the spectral functions
$G_{pp}(\o)$ and $G_{p'p'}(\o')$ \cite{dhar}.  

Note that the case $\alpha < 1$ leads to infinite thermal conductivity
while $\alpha > 1$ gives a vanishing conductivity.  Thus, depending on
the properties of the heat baths, the same wire can show
either a vanishing thermal conductivity or a diverging one. 
The usual Fourier's law would predict $J \sim 1/N$, independent of
reservoirs. Thus Fourier's law is not valid in quantum harmonic chains,
even in the presence of disorder. This breakdown of Fourier's law in
$1D$ systems has been noted in a number of earlier studies on
classical systems \cite{four} which have looked at the effects of
scattering both due to  impurities and nonlinearities. 

{\it Temperature profiles}: The local temperature of a particle can be
determined from  
its average kinetic energy , $ke_l = \la p_l^2 /(2m_l) \ra$. We get
\bea
ke_l=\f{1}{2} \int_{-\infty}^{\infty} d \o m_l \o^2 [k^2 \zh^{-1}_{l,1}(\o)
\zh^{-1}_{l,1}(-\o) I(\o) \nn \\ +  k'^2 \zh^{-1}_{l,N}(\o)
\zh^{-1}_{l,N}(-\o) I'(\o) ] 
\label{kepr}
\eea
This is straightforward to evaluate numerically for given systems and
reservoirs. Here we shall consider only the special case of heat
transmission through a perfect one-dimensional harmonic 
chain attached to one-dimensional reservoirs. For the case of 
perfect contacts ({\it i.e} $k=k'=1$) Eq.~(\ref{kepr}) simplifies (for
large $N$) to   
\bea
ke_l=\f{1}{8 \pi} \int_{-\pi}^{\pi} dk \o_k [f(\o_k)+f'(\o_k)]
\label{kepri}. 
\eea
For $T=T'$ one can verify that this corresponds to the equilibrium
kinetic energy density on an infinite chain. On the other hand if we
make the couplings weak (i.e. $k,k' << 1$ ) then we can verify that the
kinetic energy density profile for $T=T'$ corresponds to that of a
finite chain of length $N$ in thermal equilibrium ( in the quantum
case the equilibrium energy density {\it depends} on $l$ ).
  
In the non equilibrium case, at high temperatures, the local
temperature is given by $T_l=2 ke_l$ 
and from Eq.~(\ref{kepri}) we get $T_l=(T+T')/2$ which is the classical
result \cite{ried}. At low temperatures and imperfect contacts $k,k'
\neq 1$ we evaluate the 
local kinetic energy profile numerically using Eq.~(\ref{kepr}). As can be seen
in Fig.~\ref{tprof} the temperature in the bulk still has the same constant
value. At the boundaries however we see a curious feature noted
earlier by \cite{ried,zurcher}: the temperature close to the hot end is {\it
lower} than the average temperature while that at the colder end is
{\it higher} than the average. The temperature profiles at lower temperatures
are only quantitatively different. 
\vbox{
\vspace{1.5cm}
\epsfxsize=7.0cm
\epsfysize=5.0cm
\epsffile{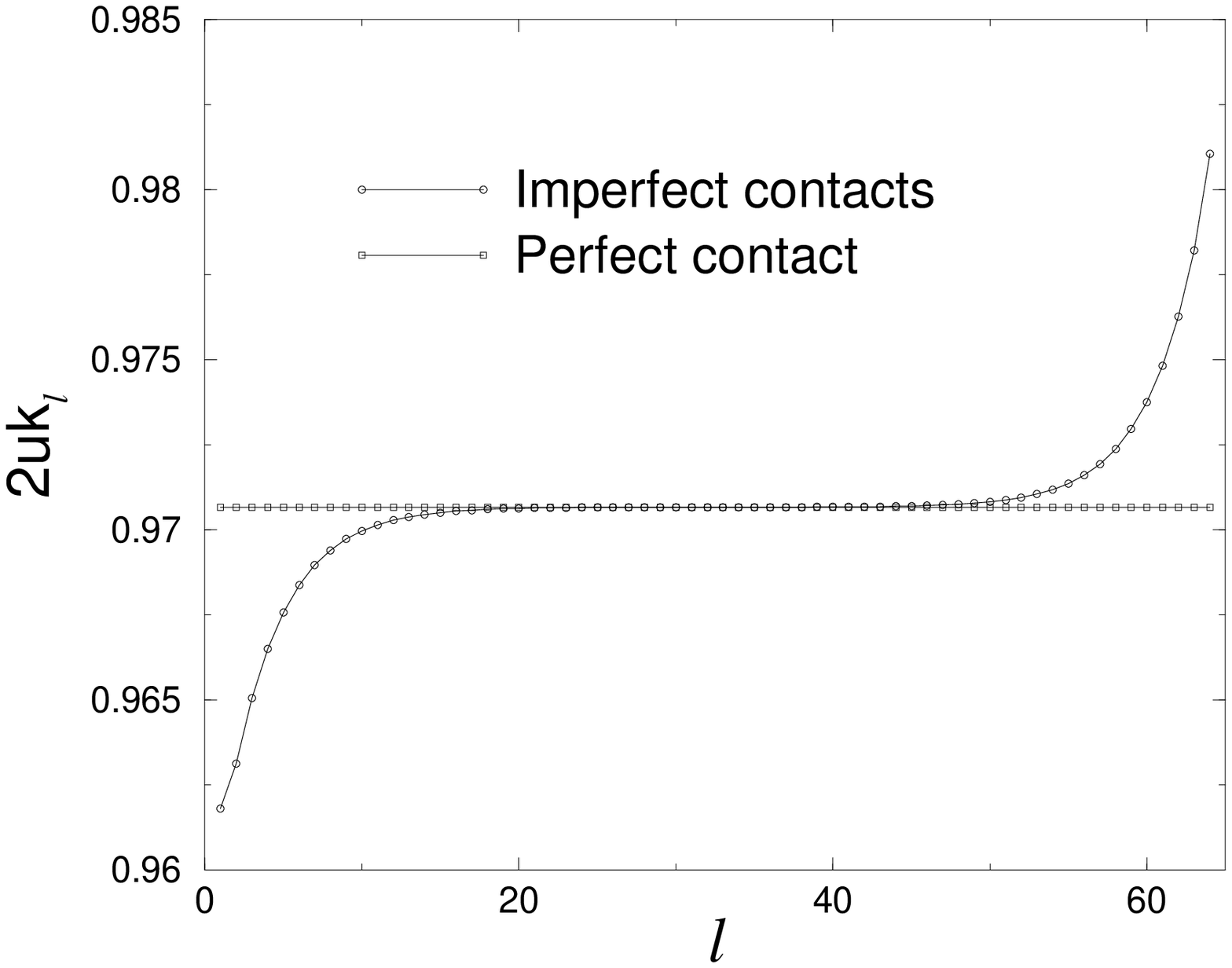}
\begin{figure}
\caption{ Kinetic energy density profile in a pure harmonic wire (N=64)
attached to one dimensional reservoirs at temperatures $T=1$ (left)
and  $T'=0.5$ (right), for perfect and imperfect contacts ($k=k'=0.9$). The
temperatures considered are quite low 
and so the bulk temperature is different from the classically expected
value $T_{av}=0.75$. 
\label{tprof}}
\end{figure}}

In summary we have used the formulation of Ford-Kac-Mazur to study
heat transport in disordered quantum harmonic chains. Exact
formal relations for the heat current have been derived and it is
shown that one obtains Landauer-like results for special reservoirs.
Earlier results on classical chains are obtained as limiting cases. 
Finally we make a couple of predictions that are interesting from the 
experimental point of view: (i) the large system size behaviour of the heat
current  is a power law and the power depends on reservoir properties
(ii) temperature profiles in perfect wires show somewhat
counterintuitive features close to contacts. A 
number of experiments on heat transport in insulating nanowires have
recently been carried out \cite{schwab}. It would be interesting to
see if our predictions, which are true for strictly one-dimensional
chains, can be verified in such experiments.     

I thank Ulrich Zurcher for pointing out errors in an earlier version
of this paper.

\end{document}